\documentclass[manuscript]{aastex62}

\usepackage[T1]{fontenc}

\usepackage{bm}        
\usepackage{amssymb, amsmath}   
\usepackage{cancel}
\usepackage{color}

\newcommand\aastex{AAS\TeX}

\DeclareMathAlphabet{\mathsfit}{\encodingdefault}{\sfdefault}{m}{sl}
\SetMathAlphabet{\mathsfit}{bold}{\encodingdefault}{\sfdefault}{bx}{sl}
\newcommand{\vect}[1]{\bm{#1}}

\DeclareMathOperator{\sech}{sech}

\received{****}
\revised{****}
\accepted{****}
\submitjournal{ApJ}

\shorttitle{\aastex\ Power-law in 3D Magnetic Reconnection}
\shortauthors{Li et al.}

\begin{document}

\title{Formation of Power-law Electron Energy Spectra in Three-dimensional
Low-$\beta$ Magnetic Reconnection}

\correspondingauthor{Xiaocan Li}
\email{xiaocanli@lanl.gov}

\author[0000-0001-5278-8029]{Xiaocan Li}
\affil{Los Alamos National Laboratory, Los Alamos, NM 87544, USA}

\author{Fan Guo}
\affiliation{Los Alamos National Laboratory, Los Alamos, NM 87544, USA}

\author{Hui Li}
\affiliation{Los Alamos National Laboratory, Los Alamos, NM 87544, USA}
\author{Adam Stanier}
\affiliation{Los Alamos National Laboratory, Los Alamos, NM 87544, USA}
\author{Patrick Kilian}
\affiliation{Los Alamos National Laboratory, Los Alamos, NM 87544, USA}

\begin{abstract}
While observations have suggested that power-law electron energy spectra are a common
outcome of strong energy release during magnetic reconnection, e.g., in solar
flares, kinetic simulations have not been able to provide definite evidence of power-laws
in energy spectra of non-relativistic reconnection. By means of 3D large-scale fully
kinetic simulations, we study the formation of power-law electron energy spectra
in non-relativistic low-$\beta$ reconnection. We find that both the global spectrum
integrated over the entire domain and local spectra within individual regions of the
reconnection layer have power-law tails with a spectral index $p \sim 4$ in the 3D
simulation, which persist throughout the non-linear reconnection phase until saturation.
In contrast, the spectrum in the 2D simulation rapidly evolves and quickly becomes soft.
We show that 3D effects such as self-generated turbulence and chaotic magnetic field
lines enable the transport of high-energy electrons across the reconnection layer and allow
them to access several main acceleration regions. This leads to
a sustained and nearly constant acceleration rate for electrons at different energies.
We construct a model that explains the observed power-law spectral index in terms of
the dynamical balance between particle acceleration and escape from main acceleration
regions, which are defined based upon a threshold for the curvature drift acceleration
term. This result could be important for explaining the formation of power-law energy
spectrum in solar flares.
\end{abstract}


\section{Introduction}
Magnetic reconnection is one of the primary mechanisms for converting magnetic
energy into plasma kinetic energy and is a major possibility for accelerating
nonthermal particles in various space, solar, and astrophysical plasmas
\citep{Zweibel2009Magnetic}. One remarkable example is solar flares, where
observations have suggested that a large amount of energetic electrons and ions
are produced during magnetic reconnection~\citep{Lin1976Nonthermal}. However,
it is still a subject of major debate on what is the resulting energy distribution
from magnetic reconnection. While there is strong observational evidence suggesting
that power-law energy distributions are a ubiquitous consequence of magnetic reconnection
in solar flare conditions~\citep{Krucker2010Measure, Krucker2014Particle,
Oka2013Kappa, Oka2015Electron, Gary2018Microwave}, this feature has not been reproduced
in self-consistent kinetic simulations in the nonrelativistic reconnection regime,
limiting our ability to study the relevant physics.

Recent kinetic simulations of magnetic reconnection in the relativistic regime
have shown the formation of power-law energy spectra~\citep[e.g.][]{Guo2014Formation,
Guo2015Particle, Guo2019Determining, Sironi2014Relativistic, Werner2014Extent}. However, obtaining
power-law distributions in the nonrelativistic regime relevant to solar flares is
considerably more difficult. Most of previous simulations were carried out with plasma
$\beta \sim 1$, with a limited amount of energy converted into plasma energy, in
comparison to the initial plasma energy~\citep{Drake2006Electron, Drake2013Power,
Dahlin2014Mechanisms, Dahlin2015Electron, Dahlin2017Role}. While 2D simulations with
low-$\beta$ condition have shown strong plasma energization, the 2D magnetic field
configuration traps high-energy particles in magnetic islands due to the
restricted particle motion across field lines~\citep{Jokipii1993Perp, Jones1998Charged},
and high-energy particle acceleration is nearly prohibited due to limited access to the
main acceleration regions--reconnection exhausts and the ends of magnetic islands
\citep{Li2015Nonthermally, Li2017Particle}. We expect that self-generated
turbulence in 3D reconnection~\citep{Bowers2007Spectral,Daughton2011Role,Dahlin2015Electron,
Dahlin2017Role,Liu2013Bifurcated,Le2018Drift,Stanier2019Influence} can mitigate this effect
and prevent particles from being trapped in magnetic islands or flux ropes, and
enable them to access multiple acceleration regions.

In this paper, we perform a 3D fully kinetic simulation
of a low-$\beta$ plasma to study the formation of power-law energy spectra in
non-relativistic reconnection. We observe nonthermal particle acceleration over an
extended time that leads to a power-law spectrum with spectral index $p\sim 4$
and about one decade in energy extent. We show that reconnection-driven turbulence enables
stronger high-energy particle acceleration by allowing particles to access several main
acceleration regions, leading to nearly constant particle acceleration rate at different
energies. In Section~\ref{sec:methods}, we describe the simulation setup and parameters.
In Section~\ref{sec:res}, we present the results on the formation of a power-law electron
energy spectrum in the 3D simulations, the transport effects in the 3D simulations, and a
simple model for the power-law index that provides an estimate consistent with PIC
simulations. In Section~\ref{sec:con}, we discuss the conclusions and implications
based on our simulation results.

\section{Numerical Simulations}
\label{sec:methods}
We carry out 2D and 3D simulations using the VPIC particle-in-cell code~\citep{Bowers2008PoP},
which solves Maxwell's equations and the relativistic Vlasov equation.
Similar to our past work~\citep{Li2015Nonthermally, Li2017Particle,Li2018Roles,Li2019Particle},
the simulations start from a force-free current sheet with $\vect{B}=B_0\tanh(z/\lambda)\vect{e}_x +
B_0\sqrt{\sech^2(z/\lambda) + b_g^2}\vect{e}_y$, where $B_0$ is the strength of the
reconnecting magnetic field, $b_g$ is the strength of the guide field $B_g$ normalized by
$B_0$, and $\lambda$ is the half-thickness of the current sheet. We choose $\lambda=d_i$ and
$b_g=0.2$ in our simulations with a mass ratio $m_i/m_e=25$, where
$d_i=c/\omega_\text{pi}=c/\sqrt{4\pi n_ie^2/m_i}$ is the ion inertial length. All simulations
have the same Alfv\'en speed $v_\text{A}\equiv B_0/\sqrt{4\pi n_0m_i}=0.2c$ and electron beta
$\beta_e\equiv8\pi nkT_e/B_0^2=0.02$. The initial particle distributions are Maxwellian with
uniform density $n_0$ and temperature $T_i=T_e=T_0$, and $kT_0=0.01m_ec^2$. Electrons are set
to have a bulk velocity drift $U_e$ so that Ampere's law is satisfied. The ratio of electron
plasma frequency and electron gyrofrequency $\omega_{pe}/\Omega_{ce}=1$.
The simulation domain is $[0 < x < L_x, -L_y/2 < y < L_y/2, -L_z/2 < z < L_z/2]$, where
$L_x=150d_i$ and $L_z=62.5d_i$ for both simulations, and $L_y=75d_i$ for the 3D simulation.
The domains are resolved using grids with $n_x\times n_z=3072\times1280$ for both simulations,
and $n_y=1536$ for the
3D simulation. We use 150 particles per species per cell. For electric and magnetic fields,
we employ periodic boundaries along the $x$- and $y$-directions and perfectly conducting
boundaries along the $z$-direction. For particles, we employ periodic boundaries along the
$x$- and $y$-directions, and reflecting boundaries along the $z$-direction. Initially,
a long wavelength perturbation with $B_z=0.02B_0$ is added to induce reconnection
\citep{Birn2001Geospace}.

\section{Results}
\label{sec:res}
\subsection{Turbulence and chaotic magnetic fields}
\begin{figure}[ht!]
  \centering
  \includegraphics[width=0.75\textwidth]{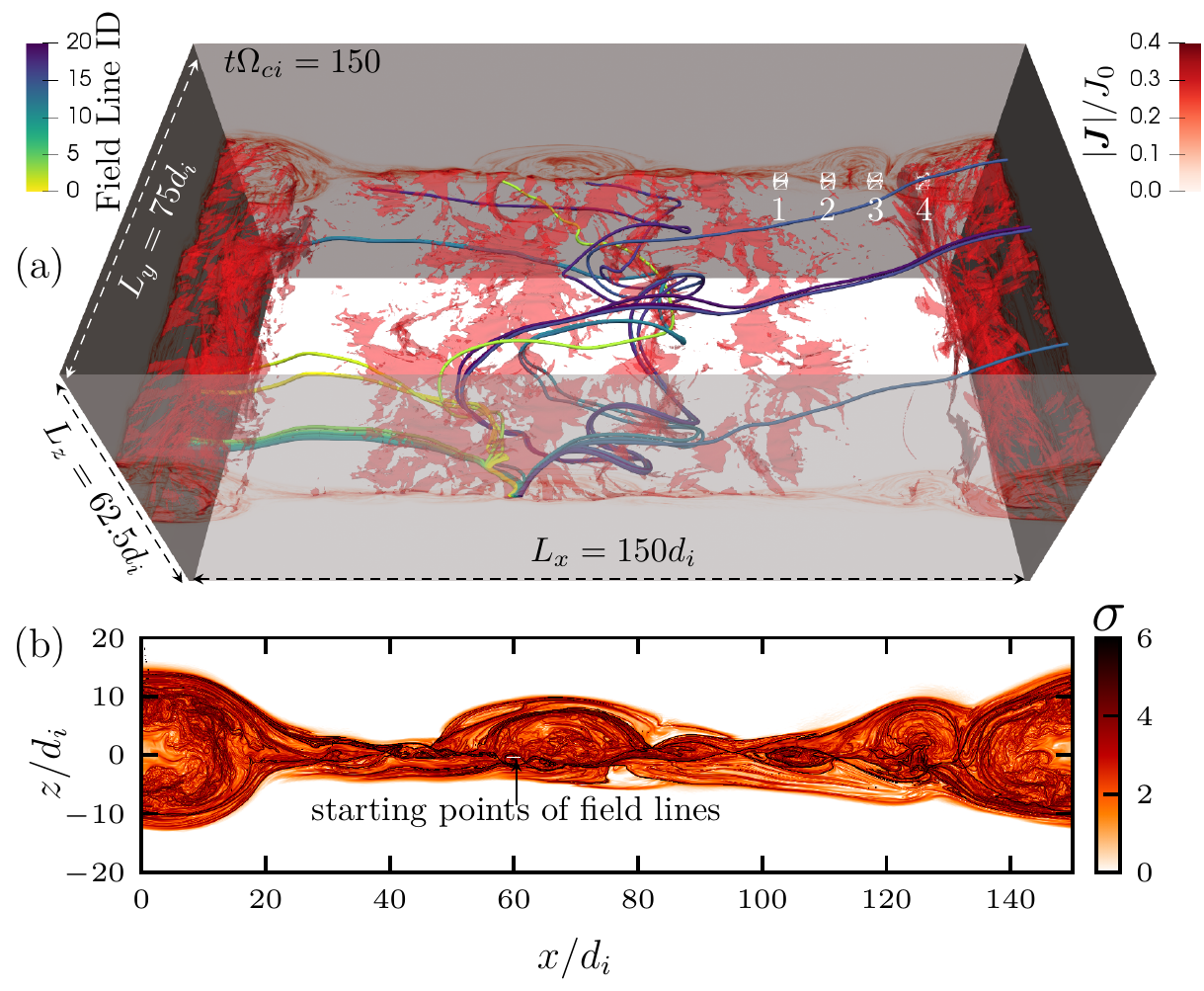}
  \caption{
    \label{fig:absJ_3d}
    Turbulence and chaotic magnetic field lines in the 3D simulation.
    (a) 3D reconnection layer at $t\Omega_{ci}=150$ showing the current density
    around the perimeter of the simulation box, an isosurface of the current density
    with $|\boldsymbol{J}|/J_0=0.3$, and magnetic field lines starting from uniformly
    distributed points along a line of $2d_i$. The field lines are color-coded with
    their seed identification numbers (IDs). The white boxes of $(2.3d_i)^3$ indicate
    regions where the local electron energy spectra are shown in Figure~\ref{fig:espect_e} (b).
    (b) Exponentiation factor $\sigma$ at $t\Omega_{ci}=150$ calculated by tracing magnetic
    field lines a distance $L_y/2$ from a plane of seed points at $y=-L_y/2$.
    The white bar indicates the starting points of the magnetic field lines shown in
    panel (a).
  }
\end{figure}

We focus on results from the 3D simulation and make comparison with 2D results where necessary.
As the reconnection proceeds, the current sheet becomes unstable to the tearing mode
instability and breaks into multiple flux ropes in the 3D simulation. These flux ropes
tend to interact and merge with each other,
and secondary flux ropes are continuously generated in the 3D reconnection layer.
These processes lead to a turbulent reconnection layer, as shown in Figure~\ref{fig:absJ_3d}
(a). At $t\Omega_{ci}=150$, three large flux ropes remain: one is in the middle of the box;
the other two near the right boundary are merging. The
isosurface of the current density shows a fragmented current layer, indicating
that turbulence is generated (see Appendix~\ref{app:turb} for the volume rendering of the
current layer and the magnetic power spectrum). Starting from 20 neighboring points along
a line of $2d_i$, the magnetic field lines quickly diverge from each other as they pass
through the fragmented current layer, which indicates that the magnetic field lines become
chaotic. To quantify this effect, we plot in Figure~\ref{fig:absJ_3d} (b) the magnetic field
line exponentiation factor $\sigma$ that measures the exponential rate of separation
of neighboring magnetic field lines~\citep{Boozer2012Separation,Daughton2014Computing,
Le2018Drift, Stanier2019Influence}. To calculate $\sigma$, we trace the magnetic field lines
a distance $L_y/2$ from a grid of points at $y=-L_y/2$ to compute the displacement map
$\vect{x}_0\rightarrow\vect{x}_f$, form the Cauchy-Green deformation tensor
$\mathcal{J}\mathcal{J}^T$ using the Jacobian of this map
$\mathcal{J}=\nabla_{\vect{x}_0}\vect{x}_f$, and calculate $\sigma$ as
$\ln(\rho_\text{max}^{1/2})$, where $\rho_\text{max}$ is the maximum eigenvalue of the
deformation tensor. Figure~\ref{fig:absJ_3d} (b) shows that $\sigma$ peaks at the
boundary regions and becomes finite inside the reconnection layer, indicating that
the magnetic field lines become chaotic. The white bar, which indicates the starting
points of the field lines shown in panel (a), crosses a boundary region with large
$\sigma$. This explains why the left part of the field lines immediately separates
from the right part in Figure~\ref{fig:absJ_3d} (a).

During these processes, about 31\% and 25\% of magnetic energy is converted into plasma
kinetic energy up to $t\Omega_{ci}=400$ in the 2D and 3D simulations, respectively.
The question is then whether the resulting particle energy spectra are different between
the two simulations.

\subsection{Electron energy spectra}
\begin{figure}[ht!]
  \centering
  \includegraphics[width=0.5\textwidth]{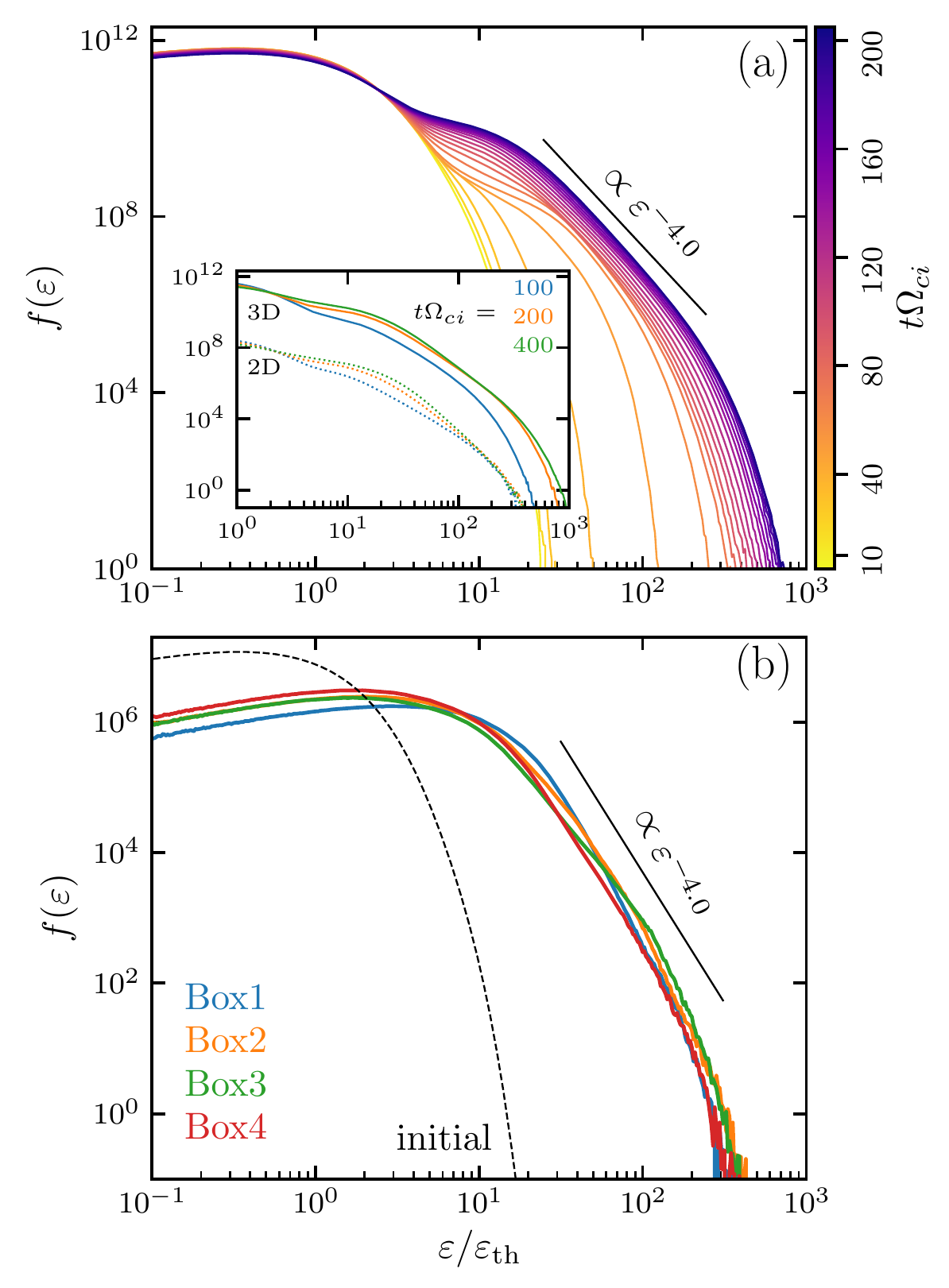}
  \caption{
    \label{fig:espect_e}
    (a) Time evolution of the global electron energy spectrum
    $f(\varepsilon)=dN(\varepsilon)/d\varepsilon$ in the 3D simulation with the embedded
    plot comparing with the 2D simulation at three time frames. $\varepsilon$ is the electron
    kinetic energy $(\gamma - 1)m_ec^2$, and $\gamma$ is the Lorentz factor. We normalize
    $\varepsilon$ by the initial thermal energy $\varepsilon_\text{th}\approx0.015m_ec^2$.
    (b) Energy spectra for electrons in the four local boxes shown in Figure
    \ref{fig:absJ_3d} (a) at $t\Omega_{ci}=150$.
  }
\end{figure}

Figure~\ref{fig:espect_e} (a) shows the time evolution of the global electron spectrum
integrated over the entire domain in the 3D simulation with the embedded plot comparing
the spectra in 2D and 3D simulations at three different time frames. The high-energy tail
($\varepsilon\in[25, 250]\varepsilon_\text{th}$) of the spectrum evolves into a power-law
$\propto\varepsilon^{-p}$ with $p \sim 4$. The power-law gradually extends to higher
energies and its spectral index does not change appreciably after $120\Omega_{ci}^{-1}$
(by 0.3 until the end of the simulation $400\Omega_{ci}^{-1}$). An additional evidence
of the nonthermal nature of the high-energy tail is that electrons are accelerated to
much higher energies (hundreds of $\varepsilon_\text{th}$) than the average free energy per
each electron-proton pair $(B_x^2/8\pi)/n\approx33\varepsilon_\text{th}$, based on the
reconnection inflow plasma parameters. Comparing to the 3D simulation,
we find that the maximum particle energy in the 2D simulation is three times smaller and
stagnates after $100\Omega_{ci}^{-1}$, which is because high-energy electrons are confined
in the magnetic islands and cannot be further energized (see Section~\ref{subsec:transport}
for more discussion). This indicates that the newly converted magnetic energy
is mostly used for accelerating low-energy electrons. Because of this, the electron flux piles up
around tens of $\varepsilon_\text{th}$ and the spectrum quickly becomes steeper. In addition,
although the spectrum in the 2D simulation appears to have a power-law tail, earlier simulations
have shown that it is actually the superposition of different thermal-like distributions in
different layers of the magnetic islands~\citep{Li2017Particle}. Because of the chaotic field
lines and self-generated turbulence in the 3D simulation, we expect that local electron
spectrum in different regions of reconnection layer to be similar. To verify this, we accumulate
energy spectra for electrons in the four regions with $(2.3d_i)^3$ each shown by white boxes
in Figure~\ref{fig:absJ_3d} (a). Regions 1--3 are in different regions of the flux rope at
$x\sim 120d_i$; region 4 is in the large flux at the boundary. In contrast to that in 2D
simulations~\citep[e.g.][]{Li2017Particle}, the local spectra shown in Figure
\ref{fig:espect_e} (b) are similar in high-energy particle flux  and the power-law high-energy tail
$\propto\varepsilon^{-4}$. This indicates efficient particle transport and mixing due to
the chaotic field lines and turbulence-induced pitch-angle scattering (see Appendix
\ref{app:anisotropy} for the low anisotropy of energetic electrons in the 3D simulation,
which indicates efficient pitch-angle scattering).

\subsection{The acceleration and transport of high-energy electrons}
\label{subsec:transport}
\begin{figure}[ht!]
  \centering
  \includegraphics[width=0.5\textwidth]{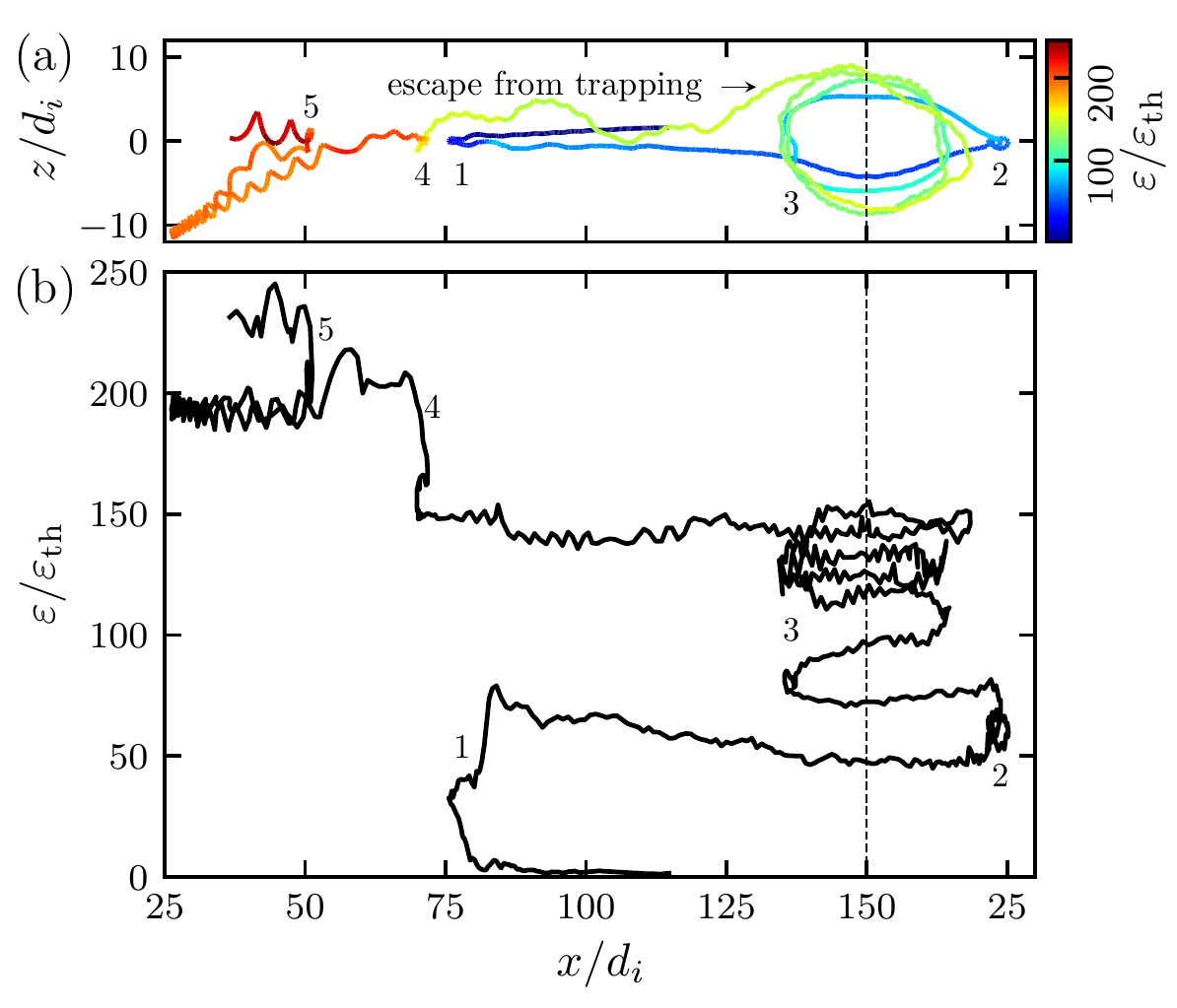}
  \caption{
    \label{fig:tracer}
    One electron trajectory in the 3D simulation. (a) The trajectory projected on the
    $x$--$z$ plane and color-coded by its kinetic energy. The numbers 1--5 indicate five
    phases of acceleration. The arrow points out when the electron escape from being
    trapped in the large flux rope. (b) $x$-position versus particle kinetic energy.
    Note that we have shifted the trajectory when electron crosses the right boundary
    at $x=150d_i$ (vertical dashed lines) to make the trajectory continuous.
  }
\end{figure}

To further demonstrate the transport effect, we traced particles as the simulations proceed
and analyze ones that are accelerated to high energy. Figure~\ref{fig:tracer}
shows one electron trajectory in the 3D simulation. Note that we have shifted the trajectory
once the electron crosses the boundary at $x=150d_i$ to make the trajectory continuous.
Figure~\ref{fig:tracer} shows that this electron sample 5 different acceleration regions.
It is first energized near the X-point at $x=75d_i$ when it enters the reconnection
region (phase 1), then streams along the magnetic field line, and gets further energized
in a small flux rope at $x=25d_i$ (phase 2). The electron is then trapped in the large flux
rope at the boundary ($x\sim150d_i$), does a typical Fermi bounce, and gets slowly energized
(phase 3). Because of the chaotic field lines and self-generated turbulence, this electron
manages to leave the flux rope, crosses the simulation domain, and gets further energized in
another exhaust region ($x\sim70d_i$) (phases 4). It is then transported to $x\sim 25d_i$,
where it is reflected by the mirror force, and gets energized to over $245\varepsilon_\text{th}$
in the exhaust region at $x\sim 50d_i$. This shows how the chaotic field lines and self-generated
turbulence in the 3D simulation enable particles to access multiple acceleration regions and
to get further accelerated, consistent with previous results~\citep{Dahlin2015Electron}.
In addition, the low plasma $\beta$ condition leads to strong particle acceleration to
hundreds of thermal energy.

Note that we choose this trajectory because of the clean separation between different
acceleration phases, which makes it good for illustrating the idea of 3D transport. Most
features described are typical in other particle trajectories, but most of them are more
chaotic. We find that only a small fraction of the escaped particles can get back to the
main acceleration regions.

\begin{figure}[ht!]
  \centering
  \includegraphics[width=\textwidth]{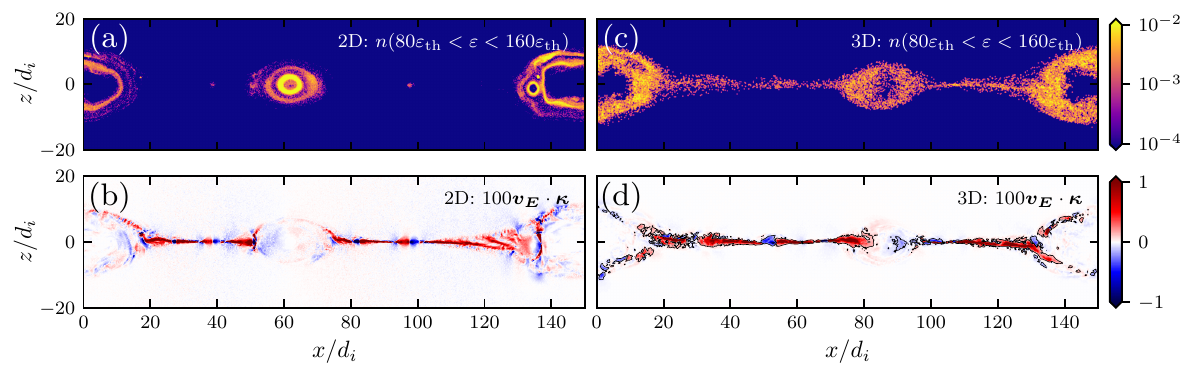}
  \caption{
    \label{fig:nhigh}
    Spatial transport of high-energy electrons. (a) The spatial distribution of
    electrons with $80\varepsilon_\text{th}<\varepsilon<160\varepsilon_\text{th}$
    at $t\Omega_{ci}=150$ in the 2D simulation. (b) $100\vect{v}_{\vect{E}}\cdot\vect{\kappa}$
    at $t\Omega_{ci}=150$ in the 2D simulation, where $\vect{v}_{\vect{E}}$ is the
    $\vect{E}\times\vect{B}$ drift velocity and $\vect{\kappa}$ is the magnetic curvature.
    (c) A $y$-slice ($y=5.5d_i$) of the spatial distribution of electrons with
    $80\varepsilon_\text{th}<\varepsilon<160\varepsilon_\text{th}$
    at $t\Omega_{ci}=150$. (d) A $y$-slice of $100\vect{v}_{\vect{E}}\cdot\vect{\kappa}$
    at the same $y$-location as panel (c).
    The black contour is at $|\vect{v}_{\vect{E}}\cdot\vect{\kappa}|=0.001$, indicating
    the boundary of the major acceleration regions.
    Note that the void at the boundary is caused by the initial perturbation
    \citep{Birn2001Geospace}.
  }
\end{figure}

Because of the enhanced spatial transport, high-energy electrons will be more broadly distributed
in the 3D simulation. To verify this, we compare the spatial distribution of the high-energy
electrons in the simulations in Figure~\ref{fig:nhigh}. In the 2D simulation, electrons with
$80\varepsilon_\text{th} < \varepsilon <160\varepsilon_\text{th}$ are confined in magnetic
islands and develop shells or rings (panel (a)) due to the restricted particle motion
across field lines~\citep{Jokipii1993Perp,Jones1998Charged}. As a result, these electrons cannot access
the reconnection exhaust regions ($x=20-50d_i$ and $75-130d_i$), where the magnetic field
is strongly bent and the energy conversion rate is the largest~\citep{Dahlin2014Mechanisms,
Li2015Nonthermally, Li2017Particle}. To clarify this, we plot
$\vect{v}_{\vect{E}}\cdot\vect{\kappa} = (\vect{B}\times\vect{\kappa}/B^2)\cdot\vect{E}$
in Figure~\ref{fig:nhigh} (b), where $\vect{v}_{\vect{E}}$ is the $\vect{E}\times\vect{B}$
drift velocity and $\vect{\kappa}=\vect{b}\cdot\nabla\vect{b}$ is the magnetic curvature
with $\vect{b}$ as a unit vector along the magnetic field. This term is proportional to
the acceleration rate associated with particle curvature drift, which is
$\varepsilon_\parallel(\vect{B}\times\vect{\kappa}/B^2)$, where $\varepsilon_\parallel$ is
the parallel kinetic energy of a particle. Figure~\ref{fig:nhigh} (b) shows that
$\vect{v}_{\vect{E}}\cdot\vect{\kappa}$ peaks at the reconnection exhaust and the two
ends of a magnetic island. In contrast, in the 3D simulation, high-energy electrons are
uniformly distributed in most regions of the flux ropes (panel (c)). More importantly, they
are transported into the reconnection exhausts ($x=25-75d_i$ and $100-125d_i$) and the two
ends of a magnetic island, so they can access the major acceleration regions similar to electrons
with lower energies (panel (d)). Therefore, we expect that these electrons should have a
nearly constant acceleration rate.

\subsection{Particle acceleration rate}
\begin{figure}[ht!]
  \centering
  \includegraphics[width=0.5\textwidth]{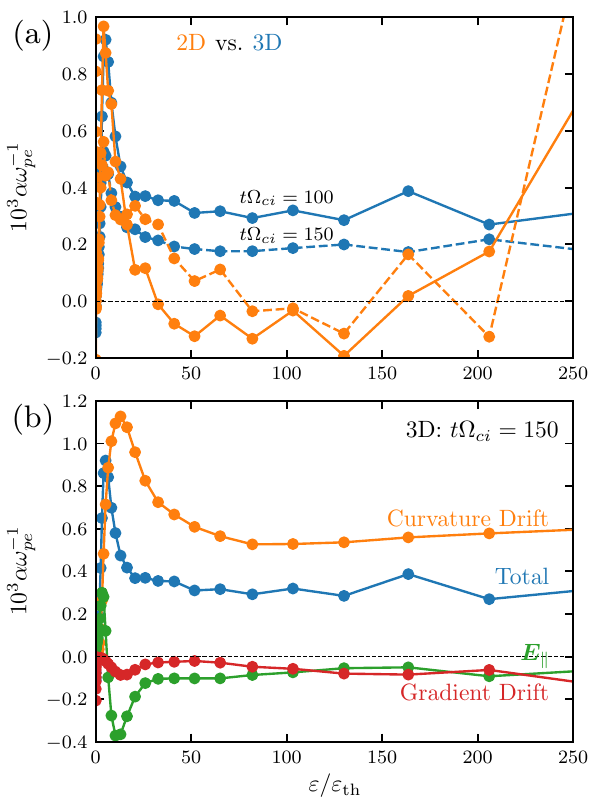}
  \caption{
    \label{fig:ene_curv_e}
    Diagnostics on the Fermi-type acceleration mechanism. (a) Electron acceleration
    rate $\alpha(\varepsilon)\equiv\left<\dot{\varepsilon}/\varepsilon\right>$
    for the 2D (orange) and 3D (blue) simulations at $t\Omega_{ci}=100$ (solid) and
    150 (dashed), where $\left<\dots\right>$ is done for electrons in different energy
    bands, $\dot{\varepsilon}=-e\vect{v}\cdot\vect{E}$, and $\vect{v}$ is the electron
    velocity. Due to the small number of high-energy electrons, $\alpha$ peaks and
    fluctuates strongly at high energies in the 2D simulation. We have run another
    2D simulation with 1500 particles/cell/species and found that the fluctuation
    level decreases and $\alpha$ at high energies is much smaller in the 2D simulation.
    (b) Electron acceleration associated with curvature drift, gradient drift,
    and the parallel electric field in the 3D simulation at $t\Omega_{ci}=150$.
  }
\end{figure}

We use all electrons (about 590 million) in the 2D simulation and 2.5\% of all
electrons (about 22.6 billion) in the 3D simulation to calculate the acceleration rate
$\alpha(\varepsilon)\equiv\left<\dot{\varepsilon}/\varepsilon\right>$,
where $\left<\dots\right>$ is the average for electrons in different energy bands,
$\dot{\varepsilon}=-e\vect{v}\cdot\vect{E}$, and $\vect{v}$ is the electron velocity.
Figure~\ref{fig:ene_curv_e} (a) shows that $\alpha$ peaks around $5\varepsilon_\text{th}$
in both 2D and 3D simulations. This is because the regions, where
the radius of magnetic curvature $|\vect{\kappa}|^{-1}\sim d_e$ (magnetic field lines are
the most strongly bent) and the acceleration rate associated with particle curvature drift
($\propto\vect{v}_{\vect{E}}\cdot\vect{\kappa}$) is the strongest, are only effective
at accelerating low-energy electrons ($\varepsilon<10\varepsilon_\text{th}$) with a
gyroradius $\leq d_e$. For the 3D simulation, $\alpha$ is nearly a constant for
$\varepsilon>40\varepsilon_\text{th}$. In contrast, $\alpha$ sharply decreases with
particle energy and even becomes negative for some energies in the 2D simulation, which
explains why the maximum energy does not change and the spectrum keeps getting steeper
after $t\Omega_{ci}=100$ in the 2D simulation. Note that $\alpha(\varepsilon)$ decreases
with time as the simulation evolves in the 3D case. This is partly because the reconnection
rate and energy conversion rate decrease, and because $\alpha(\varepsilon)$ is averaged
for all high-energy particles but most of them are in the large flux ropes where acceleration
is weak. Therefore, we need to separate particles in the major acceleration region from
that in the other regions where acceleration is weak. To accomplish this, we will distinguish
the major particle acceleration regions based on the acceleration mechanisms.

To reveal the acceleration mechanism, we evaluate betatron acceleration and decompose
$\vect{v}$ into $\vect{v}_\parallel$ that is parallel to the local magnetic field, and
the guiding-center drift velocities including curvature drift, gradient drift, inertial
drift, parallel drift, and polarization drift~\citep{Northrop1963Adabatic, LeRoux2015Kinetic, Li2019Particle}.
Figure~\ref{fig:ene_curv_e} (b) shows the three most important acceleration terms
due to the parallel electric field (or that associated with $\vect{v}_\parallel$), associated
with curvature drift, and gradient drift, respectively. Among these terms, the largest term is associated
with curvature drift, consistent with previous 2D studies based on fluid quantities
\citep{Dahlin2014Mechanisms, Li2015Nonthermally, Li2017Particle} and on particles
\citep{Li2019Particle}. Figure~\ref{fig:ene_curv_e} (b) also shows that $\vect{E}_\parallel$
accelerates thermal particles ($\sim\varepsilon_\text{th}$) but decelerates particles
with $\varepsilon>5\varepsilon_\text{th}$, and that gradient drift gives deceleration
for all particles. These results validate the assumption made in Figure~\ref{fig:nhigh}
to use $\vect{v}_{\vect{E}}\cdot\vect{\kappa}$ to separate the major acceleration
region.

\subsection{Model for spectral index}
\begin{figure}[ht!]
  \centering
  \includegraphics[width=0.5\textwidth]{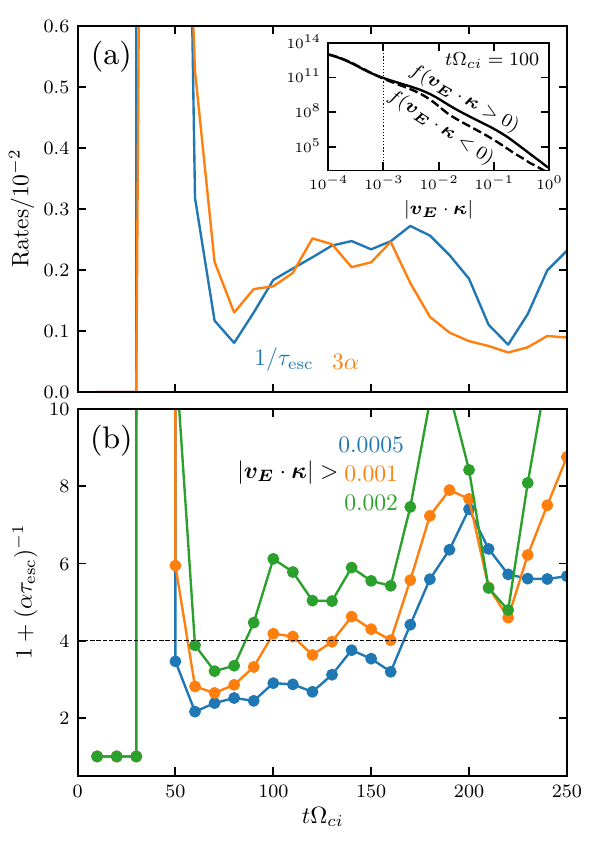}
  \caption{
    \label{fig:spect_index_e}
    An estimate of the power-law index by evaluating the acceleration rate $\alpha$
    and the escape rate $1/\tau_\text{esc}$ for high-energy electrons
    ($\varepsilon>40\varepsilon_\text{th}$) in the major acceleration region,
    where $|\vect{v}_{\vect{E}}\cdot\vect{\kappa}|$ is larger than a threshold,
    as indicated in Figure~\ref{fig:nhigh} (c). (a) Time evolution of $3\alpha$
    and $1/\tau_\text{esc}$ when $|\vect{v}_{\vect{E}}\cdot\vect{\kappa}|>0.001$.
    The embedded plot compares the distributions of the regions with negative
    and positive $\vect{v}_{\vect{E}}\cdot\vect{\kappa}$ at $t\Omega_{ci}=100$.
    The vertical dashed line indicates the chosen threshold 0.001 for
    $|\vect{v}_{\vect{E}}\cdot\vect{\kappa}|$. (b) The estimated power-law index for
    a Fermi-type acceleration mechanism $1+(\alpha\tau_\text{esc})^{-1}$ for three
    thresholds for $|\vect{v}_{\vect{E}}\cdot\vect{\kappa}|$. The dashed line indicate
    a power-law index 4, as obtained in the 3D simulation.
  }
\end{figure}

To explain the spectral index observed in the 3D simulation, we separate the main
acceleration region from the rest of the reconnection layer. Particle transport into the
non-acceleration region is simply treated as an "escape" effect.
In order to decide the criteria for the major acceleration regions, we accumulate
the PDFs of the computation cells with positive and negative
$\vect{v}_{\vect{E}}\cdot\vect{\kappa}$. The embedded plot of Figure~\ref{fig:spect_index_e}
(a) shows an example of the distributions at $t\Omega_{ci}=100$.
For $|\vect{v}_{\vect{E}}\cdot\vect{\kappa}|<0.001$, the regions with positive
$\vect{v}_{\vect{E}}\cdot\vect{\kappa}$ balances that with negative values, and the
acceleration rate for particles in these regions will be $\ll 0.001$, so these regions
do not contribute to the high-energy particle energization. Figure~\ref{fig:nhigh}
(d) shows that regions with positive $\vect{v}_{\vect{E}}\cdot\vect{\kappa}$ are usually
accompanied with regions with negative $\vect{v}_{\vect{E}}\cdot\vect{\kappa}$,
for example, near flux ropes at $x\sim55d_i$ and $x\sim90d_i$.
Therefore, we choose
$|\vect{v}_{\vect{E}}\cdot\vect{\kappa}|$ around 0.001 as the threshold for separating
the major acceleration regions and treat particles getting out these regions as escaped
particles.

After separating the major acceleration regions, we then calculate the acceleration rate
associated with curvature drift for high-energy electrons ($\varepsilon>40\varepsilon_\text{th}$)
and their escape rate $r = 1/\tau_\text{esc}=(dN_\text{esc}/dt)/N_\text{acc}$, where $N_\text{esc}$
and $N_\text{acc}$ are the number of high-energy electrons outside and inside the major
acceleration regions, respectively. $dN_\text{esc} / dt$ is the net effect of particle
escape and re-injection at the boundaries of the main acceleration regions. For single
particles (e.g. the one shown in Figure~\ref{fig:tracer}), there is a finite
possibility that escaped particles can get back into the main acceleration regions.
Statistically, more particles escape from the main acceleration regions than that are
re-injected into the main acceleration regions. Figure~\ref{fig:spect_index_e} (a) shows
an example of the calculated rates for high-energy electrons in the major acceleration region
with $|\vect{v}_{\vect{E}}\cdot\vect{\kappa}|>0.001$. Due to the small number of accelerated
particles at the beginning of the simulation, both $\alpha$ and $\tau_\text{esc}$ have a
spike as reconnection starts around $30\Omega_{ci}^{-1}$. As more particles are accelerated,
we find that $1/\tau_\text{esc}$ approaches $3\alpha$ until $150\Omega_{ci}^{-1}$, when
$\alpha$ sharply decreases due to the boundary condition. As a result, the power-law index
for a Fermi-type acceleration mechanism~\citep{Drury1983Introduction,Guo2014Formation}
$1+(\alpha\tau_\text{esc})^{-1}\approx 4$ before
$150\Omega_{ci}^{-1}$ but suddenly increases to over 6 after that, as shown in
Figure~\ref{fig:spect_index_e} (b) (orange line). The values fluctuate around 4 because
it is difficult to decide the escape boundary of the major acceleration region in such a
turbulent system. And note that the estimated power-law index is for high-energy electrons
inside the major acceleration region, while the global spectrum shown in Figure~\ref{fig:espect_e}
is for all electrons, including that in the major acceleration region and the escaped electrons.
The power-law index of the global spectrum is a dynamical balance between particle acceleration
and escape. We have tried different thresholds for $|\vect{v}_{\vect{E}}\cdot\vect{\kappa}|$.
Figure~\ref{fig:spect_index_e} (b) shows that the power-law index
increases when the threshold is higher.

\section{Discussions and Conclusions}
\label{sec:con}
By means of self-consistent kinetic simulations, we study the formation of power-law energy
spectrum in non-relativistic low-$\beta$ reconnection. We find that electrons in the 3D
simulation develop a power-law tail with a power-law index $p\sim 4$. In contrast, the
spectrum in the corresponding 2D simulation quickly becomes steeper as the simulation
proceeds. We show that the 3D effects such as self-generated turbulence and chaotic
magnetic field lines enable high-energy electrons to access several major acceleration regions,
leading to a nearly constant acceleration rate for electrons at different energies. This
enables the power-law tail to survive and extend to higher energy in the 3D simulation.
In contrast, most
high-energy electrons in the 2D simulation are slowly accelerated because they are confined
in magnetic islands and cannot access main acceleration regions. As a result, newly
converted magnetic energy is mostly used to accelerate low-energy electrons and the spectrum
quickly becomes steeper in the 2D simulation. The 3D effects also enable electrons to be
efficiently mixed, leading to nonthermal local particle distributions rather than the
thermal-like distributions trapped in different layers of a magnetic island
in 2D simulations~\citep{Li2017Particle}. Although the 3D effects have been studied
previously in terms of reconnection dynamics~\citep{Daughton2011Role, Liu2013Bifurcated,
Daughton2014Computing, Le2018Drift,  Stanier2019Influence} and electron
energization~\citep{Dahlin2015Electron,Dahlin2017Role}, for the first time, we show
that they are essential for the formation of power-law energy spectrum in non-relativistic
reconnection.

To explain the power-law index, we separate the acceleration region from non-acceleration
regions and calculate the electron acceleration rate $\alpha$ and escape rate
$r=1/\tau_\text{esc}$ for electrons inside the acceleration region.
The resulted power-law index that uses Fermi acceleration formula
$p=1+(\alpha\tau_\text{esc})^{-1}$~\citep{Drury1983Introduction,Guo2014Formation} fluctuates
around 4, consistent with the simulation result. This shows that the electron
power-law energy spectrum is a dynamical balance between acceleration and escape,
as in the classical Fermi-type acceleration processes. Several comprehensive models
have been developed for studying particle acceleration in non-relativistic reconnection
\citep{Drake2006Electron, Drake2013Power,Drake2018Comp, LeRoux2015Kinetic,
LeRoux2016Combining, LeRoux2018Self, Li2018Large, Montag2017Impact, Zank2014Particle,
Zank2015Diffusive, Zhao2018Unusual, Zhao2019Particle, Adhikari2019Role}, and they
all predict the formation of power-law energy distributions in certain regimes.
The new 3D simulations allow us to study power-laws in non-relativistic studies
and provide opportunities for testing those models. We defer this work to a future study.

The simulation boundary conditions could play a role in the formation of power-law
spectrum. The embedded plot of Figure~\ref{fig:espect_e} (a) shows a pileup of fluxes
fluxes around $20\varepsilon_\text{th}$ after $200\Omega_{ci}^{-1}$, resulting a steeper
spectrum with $p=4.35$ at $400\Omega_{ci}^{-1}$. This is likely caused by the periodic
boundary condition employed in the simulations, which terminates the acceleration of
most high-energy electrons that are in the flux rope at the boundary by slowing down
the reconnection outflows after $200\Omega_{ci}^{-1}$ (see a discussion on the effect
of the periodic boundary condition on energy conversion at the end of Appendix
\ref{app:turb}). A simulation with more realistic open boundary conditions enables
particles to escape from the boundaries and hence might lead to a steeper power-law
spectrum~\citep{Guo2014Formation}. We defer the work on simulation boundary conditions
to a future study.

We expect that the obtained power-law spectrum might change with simulation parameters. A larger
simulation domain will allow the power-law to extend to higher energies. A lower (higher) plasma
$\beta$ could make the spectrum harder (softer) by increasing (decreasing) the acceleration rate.
While we have shown here that power-law spectrum can be obtained in the low-$\beta$ reconnection
regime over the simulation time scale, our results do not rule out the possibility to generate
power-law energy spectra in high-$\beta$ reconnection. To develop a relatively well-defined
power-law spectrum (e.g. a decade in energy extent), acceleration has to be strong and/or
last for a long time. A criterion can be that $\alpha\tau_\text{inj}$ should be at least
a few~\citep{Guo2014Formation, Guo2015Particle}, where $\tau_\text{inj}$ is the particle
injection time from the reconnection inflow. Note that $\tau_\text{inj}$ is not just the
simulation time because the boundary condition will play an important role in a small-scale
simulation. Since the acceleration rate $\alpha$ is typically smaller in high-$\beta$
reconnection due to a limited amount of free magnetic energy~\citep[e.g.][]{Dahlin2017Role},
we anticipate that a much larger simulation and longer simulation time are required in order
to obtain a power-law energy spectrum in high-$\beta$ reconnection.

Our 2D simulation shows that the fluxes are piled up at tens of $\varepsilon_\text{th}$,
indicating that electrons are heated up to tens of $\varepsilon_\text{th}$.
According to~\citet{Shay2014Electron, Haggerty2015Competition}, the degree of electron
heating in reconnection scales as $0.033m_iv_A^2$, which is $2.2\varepsilon_\text{th}$
based on our simulation parameters. This is much smaller than the electron heating
in our simulations. The difference could be caused by different simulation setup. We use
a forcefree current sheet in which the plasma is uniform and they used a Harris current
sheet in which the current sheet plasma is different from the background plasma. There
are multiple X-points and magnetic islands in our simulations and there is one
X-point and occasional secondary islands in their simulations. The collapse
of the X-points and the coalescence of the islands will further accelerate electrons
and heat the plasma~\citep[e.g.][]{Drake2013Power}. To find out which factor determines
the difference, we need to perform a series of new simulations. We defer this study to
a future work.

Our simulations have a few limitations. First, we only perform simulations with
a weak guide field $0.2B_0$. In reconnection with a higher guide field, particle
acceleration rate will become smaller, the dominant electron acceleration mechanism
will change to be the parallel electric field~\citep{Dahlin2014Mechanisms,
Dahlin2016Parallel, Li2015Nonthermally, Li2017Particle, Wang2016Mechanisms}, and the
electron heating will be due to phase mixing in the strong guide-field regime in a
weakly collisional plasma~\citep{Numata2014Ion}. This will change the amplitude of
the acceleration rate and its energy dependence, which might lead to different energy
spectrum. Second, we perform the simulations with a low mass ratio $m_i/m_e=25$.
Our recent 2D simulations with different mass ratios have shown that electron
acceleration rate decreases with the mass ratio~\citep{Li2019Particle}. If this
conclusion holds in 3D simulations, we expect a steeper spectrum than that obtained
in this paper. Demonstrating this in 3D simulations with high mass ratios demands
much more computation resources than that are currently available. These problems all
require further studies in order to give quantitative predictions for the particle
energy spectrum in a large-scale reconnection layer, e.g. solar flares.

To conclude, we study the formation of power-law electron energy spectrum in
non-relativistic low-$\beta$ reconnection through performing both 2D and 3D fully
kinetic simulations. We find that both the global spectrum integrated over the
entire domain and local spectra within individual regions of the reconnection
layer have a power-law tail with a power-law index $p\sim 4$ in the 3D simulation.
In contrast, the spectrum in the 2D simulation keeps getting steeper. We show
that the self-generated turbulence and chaotic magnetic field lines in the
3D simulation enable high-energy electrons transport across the reconnection layer
enable them to access several main acceleration regions. This leads to a nearly constant
acceleration rate for electrons at different energies. To explain the power-law index,
we identify the major acceleration region where the acceleration associated with
particle curvature drift is strong, and calculate the electron acceleration rate
$\alpha$ and escape rate $r=1/\tau_\text{esc}$. The resulted power-law index
that uses Fermi acceleration formula $p=1+(\alpha\tau_\text{esc})^{-1}$ fluctuates
around 4, consistent with the simulation result. This shows that the electron
power-law energy spectrum is a dynamical balance between acceleration and escape,
as in the classical Fermi-type acceleration processes. These results could be
important for explaining the formation of power-law energy spectra in
non-relativistic plasmas, e.g. solar flares.

\acknowledgments
We thank the anomonous referee for helping to improve the manuscript.
This work was supported by NASA grant NNH16AC60I. H.L. acknowledgess the support by DOE/OFES.
F.G.'s contribution is partly based upon work supported by the U.S. Department of Energy,
Office of Fusion Energy Science, under Award Number DE-SC0018240.
We also acknowledge support by the DOE through the LDRD program at LANL. We gratefully
acknowledge our discussions with Bill Daughton, Ari Le, Xiangrong Fu, and Senbei Du.
This research used resources provided by the Los Alamos National Laboratory Institutional
Computing Program, which is supported by the U.S. Department of Energy National Nuclear
Security Administration under Contract No. 89233218CNA000001, and resources of the National
Energy Research Scientific Computing Center (NERSC), a U.S. Department of Energy Office of
Science User Facility operated under Contract No. DE-AC02-05CH11231.

\appendix
\section{Additional evidence for self-generated turbulence}
\label{app:turb}
\begin{figure}[ht!]
  \centering
  \includegraphics[width=0.5\textwidth]{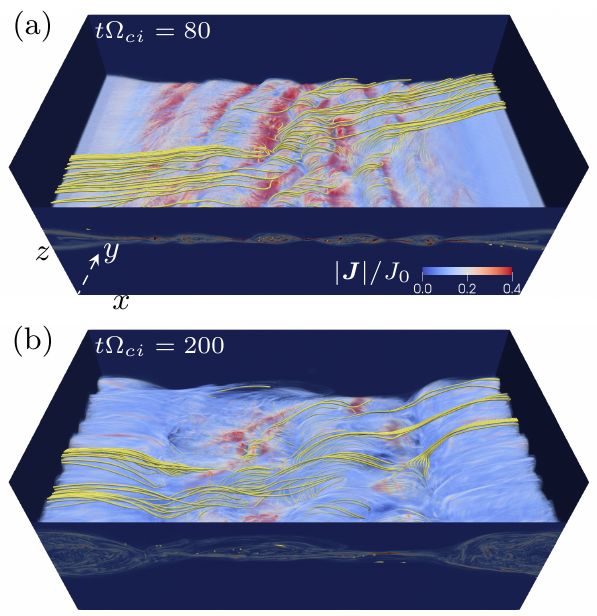}
  \caption{
    \label{fig:absj_vol}
    Volume rendering of the current density in the 3D simulation $t\Omega_{ci}=$
    (a) 80 and (b) 200. Yellow lines indicate sample magnetic field lines.
  }
\end{figure}

Figure~\ref{fig:absj_vol} shows the volume rendering of the current density in the 3D
simulation $t\Omega_{ci}=80$ and 200. At $t\Omega_{ci}=80$, the reconnection layer is
filled with flux ropes. A slice through these flux ropes shows structures that appear
to be magnetic islands, but the system is actually more complicated. These flux ropes
tend to kink and interact and merge with each other, and secondary flux ropes are
continuously generated in the layer. As the system evolves to $t\Omega_{ci}=200$,
only one large flux rope is left (besides the one at the boundary) because of the
merging of the flux ropes, and the reconnection layer becomes even more turbulent.

\begin{figure}[ht!]
  \centering
  \includegraphics[width=0.5\textwidth]{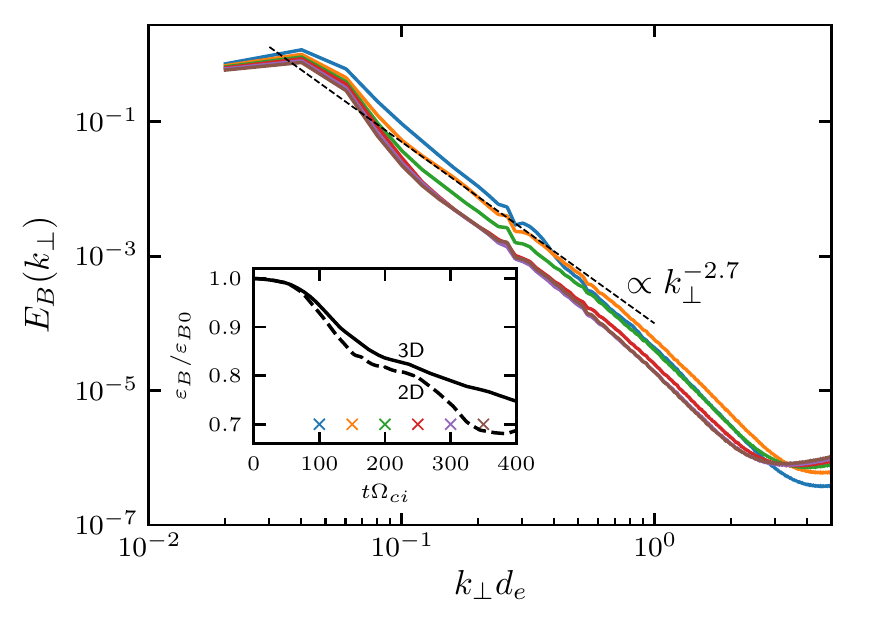}
  \caption{
    \label{fig:power_mag}
    Magnetic power spectra at five time frames indicated by the crosses in the
    embedded plot. The black dashed line indicates a power-law $\propto k_\perp^{-2.7}$.
    The embedded plot also shows the time evolution of the magnetic energy
    $\varepsilon_B$ for both simulations. $\varepsilon_{B0}$ is the initial magnetic energy.
  }
\end{figure}

We then verify the generation of turbulence by calculating the
magnetic power spectrum. We subtract $B_g=0.2B_0$ from $B_y$, apply a Blackman
window along the $z-$direction, and choose the guide-field direction as the parallel
direction. Figure~\ref{fig:power_mag} shows that the magnetic power spectrum develops
a power-law $\propto k_\perp^{-2.7}$ at large scales ($k_\perp d_e < 0.3$) after
$t\Omega_{ci}=100$ and gradually steepens at small scales.

The embedded plot in Figure~\ref{fig:power_mag} shows that the energy conversion
features two fast phases with a slow phase in between for the 2D simulation, and a
fast phase followed by a long slow phase in the 3D simulation. The fast to slow
transition occurs when the reconnection outflows collide at the periodic boundary
along the $x-$direction, which slows down the outflows and hence reduces the motional
electric field that accelerates most particles.

\section{Anisotropy of electrons at different energies}
\label{app:anisotropy}
\begin{figure}[ht!]
  \centering
  \includegraphics[width=0.5\textwidth]{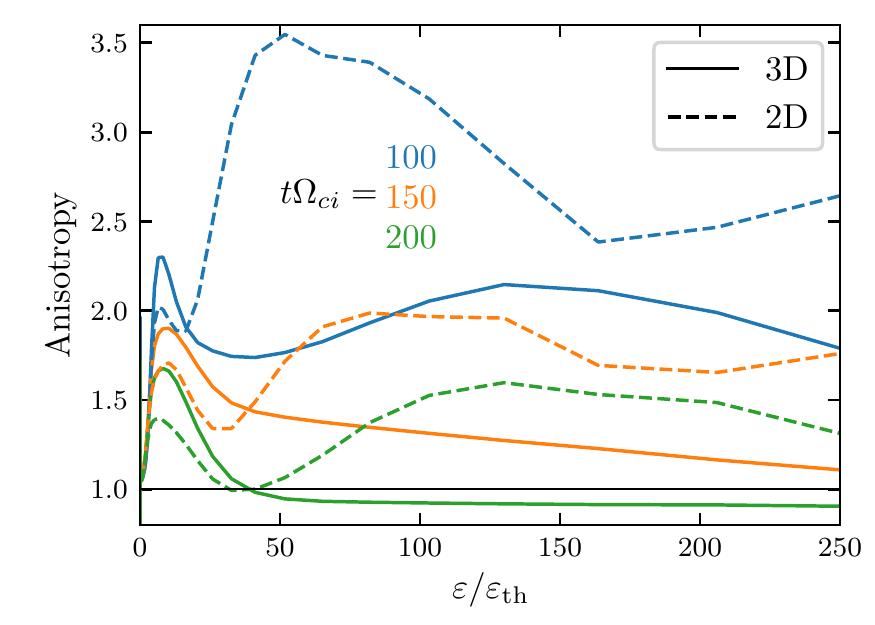}
  \caption{
    \label{fig:anisotropy}
    Anisotropy of electrons at different energies. Here the anisotropy is defined
    as $\Delta p_{e\parallel}/\Delta p_{e\perp}$ for different energy band, where
    $\Delta p_{e\parallel}=\sum(\vect{v}_\parallel-\vect{v}_{e\parallel})
    \cdot(\vect{p}_\parallel-\vect{p}_{e\parallel}/n_e)$ and
    $\Delta p_{e\perp}=0.5\sum(\vect{v}_\perp-\vect{v}_{e\perp})
    \cdot(\vect{p}_\perp-\vect{p}_{e\perp}/n_e)$ are the contributions of the electrons
    at different energy band to the parallel and perpendicular pressure, respectively.
    $\vect{v}_\parallel$ and $\vect{v}_\perp$ are the electron parallel and perpendicular
    velocity, respectively. $\vect{p}_\parallel$ and $\vect{p}_\perp$ are the electron
    parallel and perpendicular momentum, respectively. $n_e$ is the electron number
    density. $\vect{v}_{e\parallel}$ and $\vect{v}_{e\perp}$ are the electron parallel
    and perpendicular flow velocities, respectively. $\vect{p}_{e\parallel}$ and
    $\vect{p}_{e\perp}$ are the electron parallel and perpendicular momentum density,
    respectively.
  }
\end{figure}

To show the effect of pitch-angle scattering due to self-generated turbulence 3D reconnection,
we calculate the anisotropy of electrons at different energies and show the result at
3 time frames in Figure~\ref{fig:anisotropy}. As reconnection proceeds, the anisotropy
level decreases in both 2D and 3D simulations. Comparing 2D results with 3D results,
we find that the anisotropy of energetic electrons is weaker in the 3D simulation than that
in the 2D simulation. At $t\Omega_{ci}=200$, the anisotropy of energetic electrons in the 3D
simulation is close to 1. These results suggest that the self-generated turbulence in
3D reconnection can scatter energetic electrons and leads to nearly isotropic electron
distributions.

\bibliography{references}{}
\bibliographystyle{aasjournal}
\end{document}